\documentclass[10pt,conference]{IEEEtran}
\usepackage{subfigure}
\usepackage{graphicx}
\usepackage{amsmath}
\usepackage{epsfig}
\usepackage{amsfonts}
\usepackage{amssymb}
\usepackage{gensymb}
\usepackage{epstopdf}

\newcommand{\vh}{{\bf h}}

\newcommand{\vx}{{\bf x}}
\newcommand{\vy}{{\bf y}}

\newcommand{\vn}{{\bf n}}

\newcommand{\mh}{{\bf H}}

\newcommand{\se}{{\mathbb E}}

\begin{document}
\title{{\LARGE
Coded Index Modulation for Non-DC-Biased OFDM in Multiple LED 
Visible Light Communication}}
\author{
S. P. Alaka, T. Lakshmi Narasimhan$^\dagger$, and A. Chockalingam \\
Department of ECE, Indian Institute of Science, Bangalore 560012 \\
$\dagger$ Presently with National Instruments Private Limited,
Bangalore 560029
}
\maketitle
\begin{abstract}
Use of multiple light emitting diodes (LED) is an attractive way to
increase spectral efficiency in visible light communications (VLC). 
A non-DC-biased OFDM (NDC OFDM) scheme that uses two LEDs has been 
proposed in the literature recently. NDC OFDM has been shown to 
perform better than other OFDM schemes for VLC like DC-biased OFDM 
(DCO OFDM) and asymmetrically clipped OFDM (ACO OFDM) in multiple 
LEDs settings. In this paper, we propose an efficient multiple LED 
OFDM scheme for VLC which uses {\em coded index modulation}. The 
proposed scheme uses two transmitter blocks, each having a pair of 
LEDs.  Within each block, NDC OFDM signaling is done. The selection 
of which block is activated in a signaling interval is decided by
information bits (i.e., index bits). In order to improve the 
reliability of the index bits at the receiver (which is critical
because of high channel correlation in multiple LEDs settings), we 
propose to use coding on the index bits alone. We call the proposed 
scheme as CI-NDC OFDM (coded index NDC OFDM) scheme. Simulation 
results show that, for the same spectral efficiency, CI-NDC OFDM 
that uses LDPC coding on the index bits performs better than NDC OFDM.
\end{abstract}

\section{Introduction}
\label{sec1}
Optical wireless communication, where information is conveyed through 
optical radiations in free space in outdoor and indoor environments, 
is emerging as a promising complementary technology to RF wireless
communication. While communication using infrared wavelengths has been 
in existence for quite some time \cite{chann1},\cite{chann2}, more 
recent interest centers around indoor communication using visible light 
wavelengths \cite{haas1},\cite{brien}. A major attraction in indoor 
visible light communication (VLC) is the potential to simultaneously 
provide both energy-efficient lighting as well as high-speed short-range 
communication using inexpensive high-luminance light-emitting diodes (LED). 
Several other advantages including no RF radiation hazard, abundant VLC 
spectrum at no cost, and very high data rates make VLC increasingly popular. 

Orthogonal frequency division multiplexing (OFDM) which is popular in
both wired and wireless RF communications is attractive in VLC as well
\cite{ofdm1}. When OFDM is used in RF wireless communications, baseband 
OFDM signals in the complex domain are used to modulate the RF carrier.
OFDM can be applied to VLC in context of intensity modulation and
direct detection (IM/DD), where IM/DD is non-coherent and the transmit 
signal must be real and positive. This can be achieved by imposing 
Hermitian symmetry on the information symbols before the inverse fast
Fourier transform (IFFT) operation. 
Several papers have investigated OFDM in VLC \cite{ofdm1}-\cite{ofdm6}, 
which have shown that OFDM is attractive in VLC systems. A 3 Gbps 
single-LED VLC link based on OFDM has been reported in \cite{ofdm7}. 

Several techniques that generate VLC compatible OFDM signals in the 
positive real domain have been proposed in the literature 
\cite{dco2}-\cite{ndc}. These techniques include DC-biased optical 
(DCO) OFDM \cite{dco2}, asymmetrically clipped optical (ACO) OFDM
\cite{aco1}-\cite{aco3}, flip OFDM \cite{flip1},\cite{flip2}, and 
non-DC biased (NDC) OFDM \cite{ndc}. In the above works, DCO OFDM, 
ACO OFDM, and flip OFDM are studied for single-LED systems. The 
NDC OFDM in \cite{ndc} uses two LEDs. In \cite{ndc}, it has been 
that NDC OFDM performs better compared with DCO OFDM and ACO OFDM 
that use two LEDs. 

Use of multiple LEDs is a natural and attractive means to achieve 
increased spectral efficiencies in VLC. Our study in this paper 
focuses on multiple LED OFDM techniques to VLC. Our new contribution 
is the proposal of a scheme which brings in the advantage of `spatial 
indexing' to OFDM schemes for VLC. In particular, we propose a {\em `indexed
NDC (I-NDC) OFDM'} scheme, where information bits are not only conveyed
through the modulation symbols sent on the active LED, but also through
the index of the active LED. This brings in the benefit of higher rate
and better performance. Our simulation results show that, for the same
spectral efficiency, the proposed I-NDC OFDM outperforms NDC OFDM in the
low-to-moderate SNR regime. This is because, to achieve the same spectral 
efficiency, I-NDC OFDM can use a smaller-sized QAM. However, in the 
high-SNR regime, NDC OFDM performs better. We find that this is because
of the high error rates witnessed by the index bits in I-NDC OFDM due to 
high channel correlation in multiple LED settings. In order to alleviate 
this problem and improve the reliability of the index bits at the receiver, 
we propose to use coding on the index bits alone. This proposed scheme is
called {\em `coded I-NDC OFDM' (CI-NDC OFDM)} scheme. Our simulation 
results show that, for the same spectral efficiency, the proposed CI-NDC 
OFDM with LDPC coding on the index bits performs better than NDC OFDM in
VLC systems.

The remainder of this paper is organized as follows. Section. \ref{sec2}
gives an overview of DCO OFDM, ACO OFDM, flip OFDM, and NDC OFDM schemes. 
The proposed CI-NDC OFDM and performance results and discussions are 
presented in Section \ref{sec3}. Conclusions are presented in Section 
\ref{sec5}.

\begin{figure*}
\centering
\includegraphics[width=5.5in]{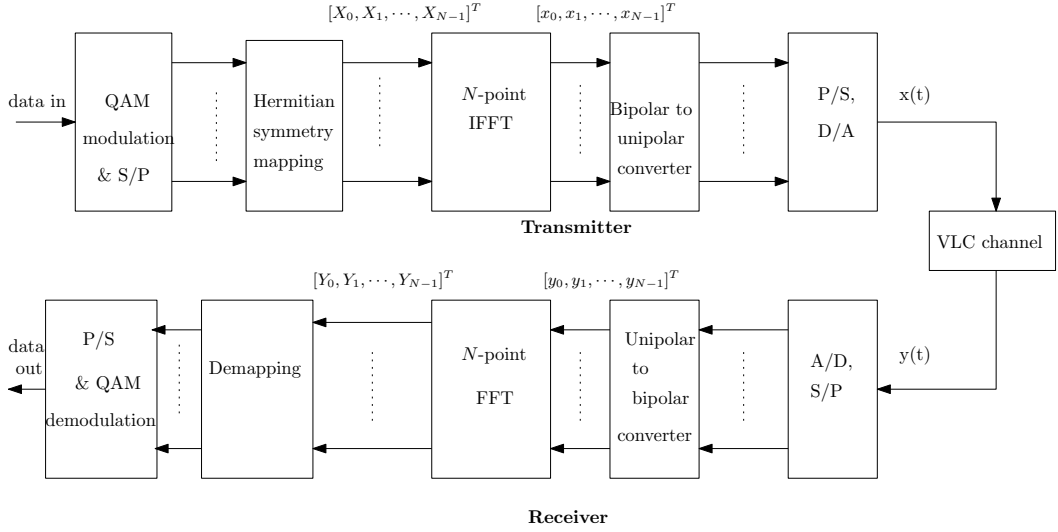}
\caption{A general single-LED OFDM system model in VLC.}
\label{ofdmsys}
\end{figure*}

\section{OFDM schemes for VLC}
\label{sec2}
Here, we present an overview of the existing OFDM schemes for VLC 
reported in the literature. Figure \ref{ofdmsys} shows the block 
diagram of a general single-LED OFDM system with $N$ subcarriers for 
VLC. In this system, a real OFDM signal is generated by constraining 
the input vector to the transmit $N$-point IFFT to have Hermitian 
symmetry, so that the output of the IFFT will be real. The output of 
the IFFT, though real, can be positive or negative. It can be made 
positive by several methods, namely, 1) adding DC bias $\big($DCO OFDM 
\cite{dco2}$\big)$, 2) clipping at zero and transmitting only positive 
part $\big($ACO OFDM \cite{aco1}-\cite{aco3}$\big)$, and 3) transmitting 
both positive and negative parts after flipping the negative part 
$\big($flip OFDM \cite{flip1}$\big)$. While the block diagram in Fig. 
\ref{ofdmsys} is for OFDM for VLC in general, the transmit and receive 
processing and achieved rates in bits per channel use (bpcu) can differ 
in the OFDM schemes listed above. These are highlighted below. 

\subsection{DCO OFDM}
\label{sec2a}
In DCO OFDM, $\big(\frac{N}{2}-1\big)\log_2M$ incoming data bits are
mapped to $\big(\frac{N}{2}-1\big)$ QAM symbols, where $M$ is the QAM
constellation size. The DC subcarrier (i.e., $X_0$) is set to zero. 
The $\big(\frac{N}{2}-1\big)$ QAM symbols are mapped to subcarriers
1 to $(\frac{N}{2}-1)$, i.e.,$\{ X_1,X_2,\cdots,X_{\frac{N}{2}-1}\}$.  
Hermitian symmetry is applied to the remaining $\frac{N}{2}$ subcarriers, 
i.e., complex conjugates of the symbols on the first $\frac{N}{2}$ 
subcarriers are mapped on the second half subcarriers in the reverse 
order, where the $\big(\frac{N}{2}+1\big)$th subcarrier is set to zero.
That is, the input to the $N$-point IFFT is given by
\[
[0,\ X_1,\ X_2,\cdots,\ X_{\frac{N}{2}-1},\ 0,\ X^*_{\frac{N}{2}-1},
\cdots,\ X^*_2,\ X^*_1]^T.
\] 
This Hermitian symmetry ensures that the IFFT output will be real and
bipolar. These bipolar OFDM symbols $x(n),\ n=0,1,\cdots,N-1$, at the 
IFFT output are converted into unipolar by adding a DC bias, $B_{dc}$.
Let $x(t)$ be the bipolar OFDM signal without DC bias. Then the unipolar 
OFDM signal $x_{dc}(t)$ that drives the transmit LED is given by
\[ 
x_{dc}(t)= x(t)+B_{dc},
\]
where $B_{dc}=k\sqrt{\mathbb E\{x^2(t)\}}$. We define this as a bias of
$10\log_{10}(k^2+1)$ dB. Note that $k=0$ corresponds to the case of no 
DC bias. For the DC bias to be not excessive, the negative going signal 
peaks must be clipped at zero. The performance of DCO OFDM depends on the 
amount of DC bias, which depends upon the size of the signal constellation 
\cite{aco2}. For example, large QAM constellations require high SNRs for
acceptable BERs, and therefore the clipping noise must be kept low, which, 
in turn, requires the DC bias to be large. As the DC bias increases, the 
required transmit power also increases. This makes the system power 
inefficient.

Due to Hermitian symmetry, the number of independent QAM symbols 
transmitted per OFDM symbol is reduced from $N$ to $\frac{N}{2}-1$. 
Thus, the achieved rate in DCO OFDM is 
\begin{equation}
\eta_{dco}=\frac{N-2}{2N}\log_2M\ \ \mbox{bpcu.}
\end{equation} 
At the receiver side, the output of the photo detector (PD), 
$y(t)$, is digitized using an analog-to-digital converter (ADC) and the 
resulting sequence, $y(n)$, is processed further. The DC bias is first 
removed and the sequence after DC bias removal is fed as input to the 
$N$-point FFT. The FFT output sequence is $[Y_0,\ Y_1,\cdots,Y_{N-1}]^T$. 
Only $\{Y_1,\ Y_2,\cdots,Y_{\frac{N}{2}-1}\}$ in the FFT output need to 
be demapped and demodulated to recover the transmit data.

\subsection{ACO OFDM}
\label{sec2b}
ACO OFDM does not use DC bias to convert the bipolar OFDM signal to unipolar. 
Instead, all negative values in the bipolar signal are clipped to zero. 
Clipping is a simpler operation in terms of implementation compared to DC 
bias. But this can introduce clipping noise. The effect of clipping noise 
can be alleviated significantly by sending data symbols only on the odd 
subcarriers. More specifically, if only the odd subcarriers are used, the 
intermodulation product terms generated due to clipping fall on the even 
subcarriers, which are ignored. While this addresses the clipping noise 
issue, the achieved data rate is compromised by a factor of two compared 
to DCO OFDM, i.e., in an $N$-subcarrier ACO OFDM scheme, only $\frac{N}{4}$ 
subcarriers are used for data transmission, whereas DCO OFDM uses 
$\frac{N}{2}-1$ subcarriers for data transmission.

In ACO OFDM, $\frac{N}{4}\log_2M$ incoming bits are mapped to 
$\frac{N}{4}$ $M$-QAM symbols. These symbols are mapped on the first 
$\frac{N}{4}$ odd subcarriers. The even subcarriers are set to zero. 
To ensure Hermitian symmetry, the complex conjugates of the symbols on 
the first $\frac{N}{2}$ subcarriers are mapped on the remaining 
subcarriers in the reverse order, i.e., the input to the $N$-point 
IFFT is given by
\[
[0,\ X_1,\ 0,\ X_2,\ 0,\cdots,\ X_{\frac{N}{4}},\ 0,\ X^*_{\frac{N}{4}},\cdots,\ 0,\ X^*_2,\ 0,\  X^*_1]^T.
\]
The real bipolar signal at the IFFT output is then converted to unipolar 
by clipping the signal at zero. Let $x(n),\ n=0,1,\cdots,N-1$, be the 
bipolar IFFT output signal. The unipolar signal is obtained as
\[
\mbox{$s$}(n)=
\begin{cases}
x(n), & \mbox{if }x(n)> 0 \\
0, & \mbox{if }x(n)\leq 0,
\end{cases}\] 
which drives the transmit LED after D/A conversion.
Since only $\frac{N}{4}$ subcarriers among the $N$ subcarriers are used
to carry data, the achieved data rate in ACO OFDM is given by
\begin{equation}
\eta_{aco}=\frac{1}{4}\log_2M\ \mbox{bpcu}.
\end{equation}
At the receiver side, the received signal $y(t)$ is first digitized to 
get $y(n), \ n=0,1,\cdots,N-1$. This sequence is input to the $N$-point 
FFT. From the $N$-point FFT output, we take only the first $\frac{N}{4}$ 
odd subcarrier data, i.e., $\{Y_1,\ Y_3,\cdots,Y_{\frac{N}{2}-1}\}$, and 
demodulate them to recover the transmit data.

\subsection{Flip OFDM}
\label{sec2c}
Flip OFDM is similar to DCO OFDM except DC biasing. Instead of DC biasing, 
it uses two OFDM symbols to send the bipolar signals, i.e., positive and 
negative parts are sent as two consecutive OFDM symbols. Like in DCO OFDM, 
in flip OFDM also, the input to the $N$-point IFFT is 
\[
[0,\ X_1,\ X_2,\cdots,\ X_{\frac{N}{2}-1},\ 0,\ X^*_{\frac{N}{2}-1},\cdots,\ X^*_2,\ X^*_1]^T.
\] 
Let $x(n),\ n=0,1,\cdots,N-1,$ be the bipolar $N$-point IFFT output. The 
IFFT output $x(n)$ is fed to a polarity separator, which separates the 
positive and negative parts of $x(n)$. That is, the sequence $x(n)$ can 
be written in the form
\[
x(n)=x^{+}(n)+x^{-}(n),
\]
where
\begin{eqnarray}
\mbox{$x^+$}(n)=
\begin{cases}
x(n), & \mbox{if }x(n)\geq 0 \\
0, & \mbox{if }x(n)< 0
\end{cases}, \\
\mbox{$x^-$}(n)=
\begin{cases}
x(n), & \mbox{if }x(n)< 0 \\
0, & \mbox{if }x(n)\geq 0.
\end{cases}
\end{eqnarray}
The positive part $x^{+}(n)$ is transmitted as the first OFDM symbol. 
The polarity inverted (i.e., flipped) negative part $\big($i.e., 
$-x^{-}(n)\big)$ is transmitted as the second OFDM symbol. Note that 
the positive and negative parts of the bipolar OFDM signal are 
transmitted as two consecutive OFDM symbols with the negative part 
flipped. Since $\big(\frac{N}{2}-1\big)$ $M$-QAM symbols are sent in 
two slots, the achieved data rate in flip OFDM is 
\begin{eqnarray}
\eta_{flip} & = & \ \frac{\frac{N}{2}-1}{2N}\log_2M \nonumber \\
& \approx & \frac{1}{4}\log_2M\mbox{bpcu, \hspace{2mm} for large }N.
\label{etaflip}
\end{eqnarray}
At the receiver, the received signal $y(t)$ is first digitized. Let 
$y^+(n)$ and $y^-(n)$ represent the time samples belonging to the first 
and second OFDM symbols, respectively. These two sample sequences are 
added; the polarity of the $y^-(n)$ is inverted before adding. Thus, the 
resulting bipolar signal $y(n)$ is given by $y(n)=y^+(n)-y^-(n)$,
which is S/P converted and fed to the $N$-point FFT. The FFT output 
values of subcarriers 1 to $\frac{N}{2}-1$ are demapped and demodulated 
to recover the transmit data.

\subsection{NDC OFDM}
\label{sec2d}
\begin{figure*}
\centering
\includegraphics[width=5.0in]{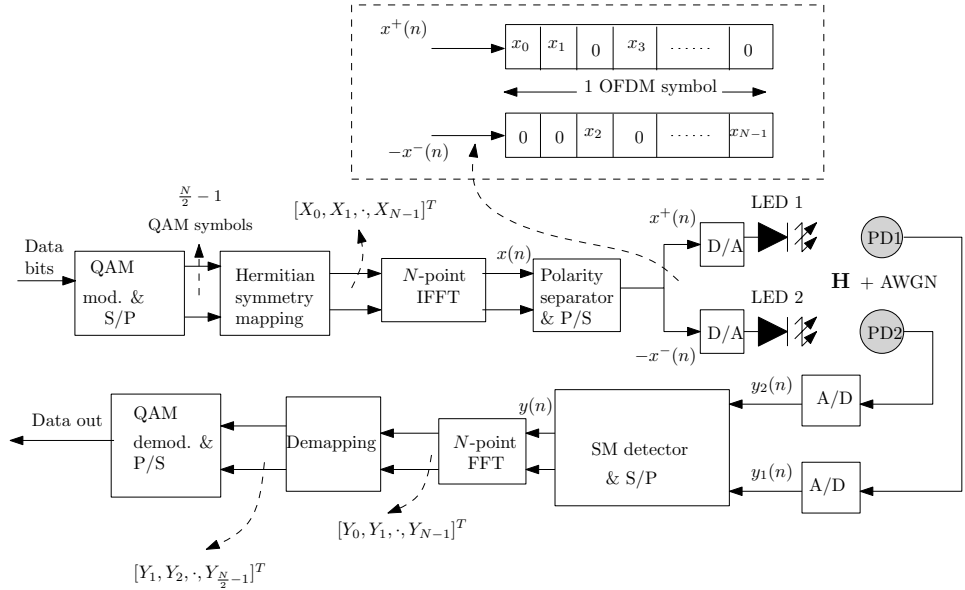}
\caption{Non-DC biased OFDM.}
\label{ndc}
\end{figure*}

NDC OFDM is similar to flip OFDM except the number of time slots used. 
Instead of sending the OFDM symbol in two consecutive time slots, this 
scheme exploits the spatial dimension. That is, this scheme uses two LEDs 
to send the bipolar signals; positive and negative parts drive two
different LEDs. Figure \ref{ndc} shows the block diagram of NDC OFDM.
As in flip OFDM, the input to the $N$-point IFFT is given by
\[
[0,\ X_1,\ X_2,\cdots,\ X_{\frac{N}{2}-1},\ 0,\ X^*_{\frac{N}{2}-1},\cdots,\ X^*_2,\ X^*_1]^T.
\] 
Let $x(n)$ be the bipolar IFFT output. As in flip OFDM, this output is 
fed to a polarity separator, which separates the positive and negative 
parts of $x(n)$, i.e., $x(n)=x^+(n)+x^-(n)$, where $x^+(n)$ and $x^-(n)$ 
are as defined in flip OFDM. $x^+(n)$ drives the first LED, and $-x^-(n)$ 
(i.e., flipped or polarity inverted signal) drives the second LED. 
Therefore, at a given time, only one LED will be active, where the index of
the active LED (i.e., LED1 and LED2) is decided by the sign of the OFDM 
signal. This scheme can be viewed as OFDM with spatial modulation (SM), 
where the LED to activate in a given channel use is chosen based on the sign.

Due to Hermitian symmetry, the number of independent QAM symbols 
transmitted per OFDM symbol is reduced from $N$ to $\frac{N}{2}-1$. 
Thus, the achieved data rate of NDC OFDM is 
\begin{equation}
\eta_{ndc}=\frac{N-2}{2N}\ \log_2M\ \mbox{ bpcu},
\label{etandc}
\end{equation}
which is the same as that of DCO OFDM.

The unipolar OFDM signal is transmitted over the VLC MIMO channel $\mh$, 
where $\mh$ is a $N_r\times N_t$ channel matrix, $N_t$ is the number of 
LEDs, and $N_r$ is the number of PDs. Here, $N_t=N_r=2$. The output of
the PDs are fed to the ADCs. The digitized output of the ADCs, denoted 
by $\vy=[y_1(n)\ \ y_2(n)]^T$, is fed as input to the SM detector. The 
SM detector, for example, can be zero forcing (ZF) detector. That is, the 
SM detector output, denoted by $y(n),\ n=0,1,\cdots,N-1$, is 
\begin{eqnarray}
|y(n)| &=& \max_{i=1,2} \ |z_i(n)|,\\
\mbox{sign}\{y(n)\} &=&
\begin{cases}
+\mbox{ve}, \mbox{ if }\displaystyle\arg\max_{i=1,2}\ |z_i(n)|=1\\
-\mbox{ve}, \mbox{ if }\displaystyle\arg\max_{i=1,2}\ |z_i(n)|=2,
\end{cases}
\end{eqnarray} 
where
\begin{eqnarray}
\begin{bmatrix} z_1(n)\\[0.25em]
z_2(n)
\end{bmatrix}
 &=& \begin{bmatrix} \big(\vh_1^{T}\vh_1\big)^{-1}\vh_1^{T}\ \vy \\[0.25em]
\big(\vh_2^{T}\vh_2\big)^{-1}\vh_2^{T}\ \vy
\end{bmatrix},
\end{eqnarray}
and $\vh_i$ is the $i$th column of channel matrix $\mh, i=1,2$.
The SM detector output $y(n)$ is then fed to the $N$-point FFT. From the 
$N$-point FFT output, the subcarriers 1 to $\big(\frac{N}{2}-1\big)$ are 
demodulated to get back the transmit data.

\subsection{DCO/ACO/Flip/NDC OFDM performance comparison}
\label{sec2e}
\begin{figure}
\centering
\includegraphics[height=1.2in]{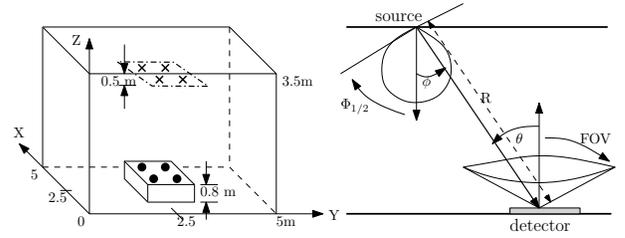}
\caption{Geometric set-up of the considered indoor VLC system.
A dot represents a photo detector and a cross represents an LED.}
\label{sys}
\end{figure}

Here, we illustrate a BER performance comparison between DCO OFDM, ACO
OFDM, flip OFDM, and NDC OFDM. The indoor VLC system set up is shown in
Fig. \ref{sys}. The system parameters of the indoor VLC system considered 
in the simulation are given in Table \ref{tab1}. All systems use $N_t=2$ 
LEDs, $N_r=2$ PDs. The PDs are kept symmetrical on top of a table with 
respect to the center of the floor with a $d_{rx}$ of 0.1m. The LEDs are 
kept symmetrical with respect to the center of the room at 1m apart and 
at 3m height(i.e., $d_{tx}=1$m and $z=3$m). The channel gain between 
$j$th LED and $i$th PD is calculated as \cite{chann2}
\begin{equation}
{h_{ij}} = \frac{n+1}{2\pi}\cos^{n}{\phi_{ij}}\,
\cos{\theta_{ij}}\frac{A}{R_{ij}^2}
\mbox{rect}\Big(\frac{\theta_{ij}}{FOV}\Big),
\label{channel}
\end{equation}
where $\phi_{ij}$ is the angle of emergence with respect to the $j$th
source (LED) and the normal at the source, $n$ is the mode number of the 
radiating lobe given by
$n=\frac{-\ln(2)}{\ln\cos{\Phi_{\frac{1}{2}}}}$,
$\Phi_\frac{1}{2}$ is the half-power semiangle of the LED \cite{new1},
$\theta_{ij}$ is the angle of incidence at the $i$th photo detector,
$A$ is the area of the detector, $R_{ij}$ is the distance between the
$j$th source and the $i$th detector, FOV is the field of view of the
detector, and $\mbox{rect}(x)=1$ if $|x|\leq 1$ and $\mbox{rect}(x)=0$ 
if $|x|> 1$. 

\begin{table}
\centering
\begin{tabular}{|l||l|l|}
\hline
        & Length $(X)$  & 5m \\ \cline{2-3}
Room    & Width ($Y$)   & 5m \\ \cline{2-3}
        & Height ($Z$)  & 3.5m  \\ \hline \hline
        & Height from the floor & 3m \\ \cline{2-3}
        & Elevation     & $-90\degree$ \\ \cline{2-3}
Transmitter & Azimuth   & $0\degree$ \\ \cline{2-3}
        & $\Phi_{1/2}$  & $60\degree$ \\ \cline{2-3}
        & Mode number, $n$ & 1 \\ \cline{2-3}
        & $d_{tx}$      & 1m \\ \hline \hline
        & Height from the floor & 0.8m\\ \cline{2-3}
        & Elevation     & $90\degree$ \\ \cline{2-3}
Receiver & Azimuth      & $0\degree$ \\ \cline{2-3}
        & Responsivity, $r$ & 1 Ampere/Watt  \\ \cline{2-3}
        & FOV           & $85\degree$ \\ \cline{2-3}
        & $d_{rx}$      & 0.1m \\ \hline
\end{tabular}
\vspace{3mm}
\caption{\label{tab1} System parameters in the considered indoor VLC system.}
\end{table}

\begin{figure}
\centering
\includegraphics[width=3.35in,height=2.5in]{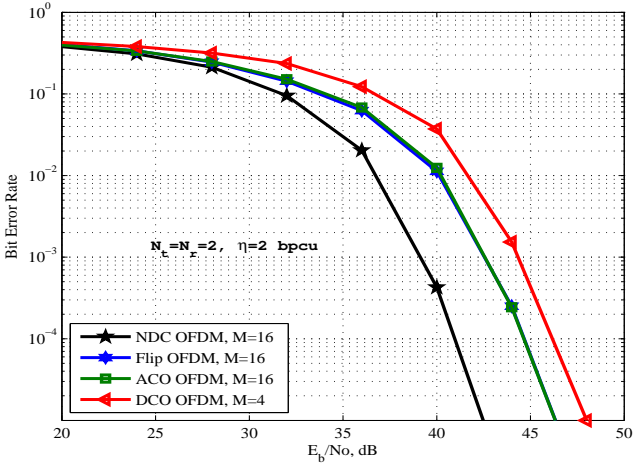}
\vspace{-2mm}
\caption{Comparison of the BER performance of DCO OFDM, ACO OFDM, 
flip OFDM, and NDC OFDM $\eta=2$ bpcu, $N_t=N_r=2$.}
\label{fig4}
\vspace{-4mm}
\end{figure}

Figure \ref{fig4} shows the BER performance achieved by DCO OFDM, ACO OFDM, 
flip OFDM, and NDC OFDM for $\eta=2$ bpcu, and $N_t=N_r=2$. The parameters 
considered these systems are: 1) DCO OFDM: $N_t=N_r=2$, $M=4$, 7 dB bias,
2) ACO OFDM: $N_t=N_r=2$, $M=16$, 3) flip OFDM: $N_t=N_r=2$, $M=16$, and 
4) NDC OFDM: $N_t=N_r=2$, $M=16$. In ACO OFDM, flip OFDM, and DCO OFDM, 
there are two parallel transmitting OFDM blocks, each drives one LED 
simultaneously. ZF detection is used for DCO OFDM, ACO OFDM, and flip
OFDM. The hypothesis testing based detection method presented in Sec. 
\ref{sec2d} is used for NDC OFDM. From Fig. \ref{fig4}, it can be seen 
that DCO OFDM has poor performance compared to other systems, and this 
is due to the DC over-biasing. Also, ACO OFDM and flip OFDM have the same 
performance. Among the OFDM schemes discussed above, NDC OFDM achieves 
better performance compared to other OFDM schemes. This is because of 
the spatial interference experienced by the other OFDM schemes, i.e.,
while two LEDs are active simultaneously in DCO OFDM, ACO OFDM, and 
flip OFDM, only one LED will be active at a time in NDC OFDM. 

\section{Proposed CI-NDC OFDM for VLC}
\label{sec3}
Motivated by the advantages of multiple LEDs and spatial indexing
to achieve increased spectral efficiency, here we first propose a 
multiple LED OFDM scheme called `indexed NDC OFDM (I-NDC OFDM)'. In
this scheme, additional bits are conveyed through the index of the
active LED. Then, realizing the need to protect the index bits better 
in this scheme, we propose to use coding on the index bits. This 
scheme is called `coded index NDC OFDM (CI-NDC OFDM). 

\subsection{Proposed I-NDC OFDM}
\label{sec3a}
\begin{figure*}
\centering
\includegraphics[width=5.0in]{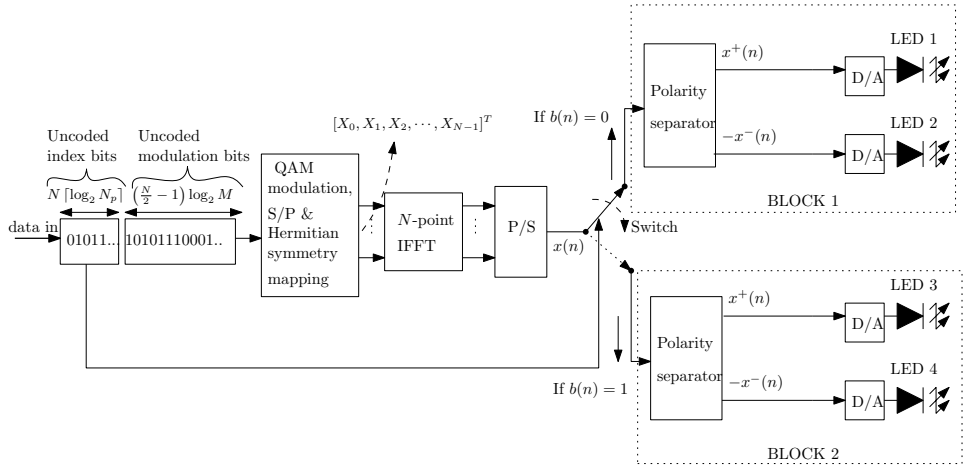}
\caption{Proposed I-NDC OFDM transmitter.}
\label{indctx}
\end{figure*}

The block diagram of the proposed I-NDC OFDM transmitter is illustrated in 
Fig. \ref{indctx}. I-NDC OFDM is an $N$-subcarrier OFDM system with $N_p$ 
pairs of LEDs and $N_r$ photo detectors, where the total number of LEDs 
$N_t=2N_p$. We consider $N_p=2$, i.e., there are 2 pairs of LEDs.
In Fig. \ref{indctx}, the $\{LED1, LED2\}$ pair forms BLOCK 1 and the
$\{LED3, LED4\}$ pair forms BLOCK 2. In each channel use, only one LED 
in either BLOCK 1 or BLOCK 2 will be activated. The choice of which BLOCK 
has to be activated in a given channel use is made based on indexing. In 
a general setting, $m$ index bits can select one BLOCK among $2^m$ BLOCKs.
In the considered system, $m=1$ and $N_p=2$. Therefore, the BLOCK selection 
is done using one index bit per channel use. The LED pair in the selected 
BLOCK will be driven as per the standard NDC OFDM scheme described in 
Sec. \ref{sec2d}. The I-NDC OFDM transmitter operation is described below.

\begin{figure*}
\centering
\includegraphics[width=4.5in]{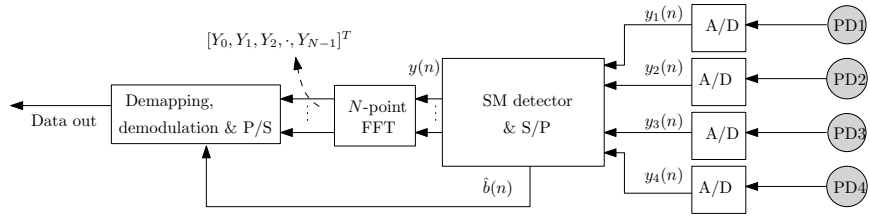}
\caption{Proposed I-NDC OFDM receiver.}
\label{indcrx}
\end{figure*}

{\em Transmitter:} As in NDC OFDM, in I-NDC OFDM also, 
$\big(\frac{N}{2}-1\big)\log_2M$ incoming data bits are first mapped 
to $\big(\frac{N}{2}-1\big)$ QAM symbols, and the input to the $N$-point 
IFFT is given by
\[
[0,\ X_1,\ X_2,\cdots,\ X_{\frac{N}{2}-1},\ 0,\ X^*_{\frac{N}{2}-1},
\cdots,\ X^*_2,\ X^*_1]^T.
\] 
This ensures real and bipolar IFFT output. Let $x(n),\ n=0,1,\cdots,N-1$,
be the IFFT output. For large $N$, (e.g., $N\geq 64$), $x(n)$ can be 
approximated as i.i.d.  real Gaussian with zero mean and variance 
$\sigma_x^2$. Therefore, $|x(n)|$ has an approximately  half-normal 
distribution with mean $\frac{\sigma_x}{\sqrt{2\pi}}$ and variance
$\frac{\sigma_x^2(\pi -2)}{2\pi}$ \cite{aco1}. The IFFT output sequence 
$x(n)$ is input to a BLOCK selector switch. For each $n,\ n=0,1,\cdots,N-1$,
the switch decides the BLOCK to which $x(n)$ has to be sent. This BLOCK
selection in a given channel use is done using index bits. Let $b(n)$
denote the index bit for the $n$th channel use, $n=0,1,\cdots,N-1$. The 
BLOCK selector switch performs the following operation:
\begin{eqnarray}
\mbox{if }b(n) &=& 0,\ \ x(n)\mbox{ goes to BLOCK1}\nonumber \\\nonumber
\mbox{if }b(n) &=& 1,\ \ x(n)\mbox{ goes to BLOCK2}.
\end{eqnarray}
In the selected BLOCK, the polarity separator separates positive and negative
parts of $x(n)$; $x(n)$ can be written as
\[
x(n)= x^+(n)+x^-(n),
\]
{\small
$\mbox{$x^+(n)$}=
\begin{cases}
x(n), & \mbox{if }x(n)\geq 0 \\
0, & \mbox{if }x(n)< 0,
\end{cases}$
$\mbox{$x^-(n)$}=
\begin{cases}
x(n), & \mbox{if }x(n)< 0 \\
0, & \mbox{if }x(n)\geq 0.
\end{cases}$
}

\vspace{2mm}
\hspace{-5mm}
If the selected BLOCK is BLOCK 1, $x^+(n)$ drives LED1 and $-x^-(n)$ 
drives LED2. Similarly, if BLOCK 2 is selected, $x^+(n)$ drives LED3 and 
$-x^-(n)$ drives LED4. So, the light intensity emitted by each LED is 
either $|x(n)|$ or 0. Since
$|x(n)|\sim\mathcal{N}_+\big(\frac{\sigma_x}{\sqrt{2\pi}},\frac{\sigma_x^2(\pi -2)}{2\pi}\big)$,
the intensity $I$ is such that
\[
I\in\bigg\{\mathcal{N}_+\bigg{(}\frac{\sigma_x}{\sqrt{2\pi}},\frac{\sigma_x^2(\pi -2)}{2\pi}\bigg{)} \bigg\}.
\]
{\emph Achieved data rate:} In I-NDC OFDM, $\big(\frac{N}{2}-1\big)$
QAM symbols are sent per OFDM symbol. In addition, 
$\left\lceil\log_2N_p\right\rceil$ number of bits are used to select 
the active BLOCK per channel use. Therefore, the achieved data rate in 
I-NDC OFDM is 
\begin{eqnarray}
\eta_{indc} &=& \bigg(\frac{\frac{N}{2}-1}{N}\bigg)\log_2M\ + \ \frac{N\left\lceil\log_2N_p\right\rceil}{N}\nonumber\\
&=& \underbrace{\bigg(\frac{N-2}{2N}\bigg)\log_2M}\ + \underbrace{\ \left\lceil\log_2N_p\right\rceil\ }\mbox{ bpcu}.\\\nonumber
& & \mbox{modulation bits}\hspace{.35in}\mbox{index bits}
\label{etaindc}
\end{eqnarray}
{\em Receiver:} The block diagram of I-NDC receiver is illustrated
in Fig. \ref{indcrx}. We assume perfect channel state information at
the receiver. Assuming perfect synchronization, the $N_r\times 1$
received signal vector
at the receiver is given by
\begin{eqnarray}
\vy = r\mh\vx+\vn,
\label{sysmodel}
\end{eqnarray}
where $\vx$ is the
$N_t\times 1$ transmit vector, $r$ is the responsivity of the detector, and
$\vn$ is the noise vector of dimension $N_r\times 1$. Each element in the
noise vector $\vn$ can be modeled as i.i.d. real AWGN with zero mean and
variance $\sigma^2$. Note that the transmit vector $\vx$ has only one 
non-zero element, and the remaining $N_t-1$ elements are zeros. The 
non-zero element in $\vx$ represents the light intensity $I$ emitted by 
the active LED, where 
$I\sim\mathcal{N}_+\big(\frac{\sigma_x}{\sqrt{2\pi}},\frac{\sigma_x^2(\pi -2)}{2\pi}\big)$.
The average received signal-to-noise ratio (SNR) is given by
${\overline{\gamma}}=\frac{r^2P_r^2}{\sigma^2}$,
where
\begin{eqnarray}
P_r^2 \ = \ \frac{1}{N_r}\sum_{i=1}^{N_r}\se [{|\mh_i\vx|}^2] 
\ = \ \frac{\sigma^2_x}{2N_r}\sum_{i=1}^{N_r}\sum_{j=1}^{N_t}h_{ij}^2,
\end{eqnarray}
and $\mh_i$ is the $i$th row of $\mh$. The received optical signals are 
converted to electrical signals by the PDs. The output of these PDs are 
then fed to the ADCs. The output of the ADCs is given by the vector 
$\vy=[y_1(n)\ \ y_2(n)\ \ y_3(n)\ \ y_4(n)]^T$, which is fed to the SM 
detector. The bipolar output of the SM detector is fed to the $N$-point 
FFT. The SM detector can be a ZF detector. That is, the SM detector output, 
denoted by $y(n),\ n=0,1,\cdots,N-1$, is 
\begin{eqnarray}
|y(n)| &=& \max_{i=1,2,3,4} \ |z_i(n)|\\
\mbox{sign}\{y(n)\} &=&
\begin{cases}
+\mbox{ve}, \mbox{ if }\displaystyle\arg\max_{i=1,2,3,4}\ |z_i(n)|=1\\
-\mbox{ve}, \mbox{ if }\displaystyle\arg\max_{i=1,2,3,4}\ |z_i(n)|=2\\
+\mbox{ve}, \mbox{ if }\displaystyle\arg\max_{i=1,2,3,4}\ |z_i(n)|=3\\
-\mbox{ve}, \mbox{ if }\displaystyle\arg\max_{i=1,2,3,4}\ |z_i(n)|=4,
\end{cases}
\end{eqnarray} where
\begin{eqnarray}
\begin{bmatrix} z_1(n)\\[0.1em]
z_2(n)\\[0.1em]
z_3(n)\\[0.1em]
z_4(n)
\end{bmatrix}
 &=& \begin{bmatrix} \big(\vh_1^{T}\vh_1\big)^{-1}\vh_1^{T}\ \vy \\[0.1em]
\big(\vh_2^{T}\vh_2\big)^{-1}\vh_2^{T}\ \vy\\[0.1em]
\big(\vh_3^{T}\vh_3\big)^{-1}\vh_3^{T}\ \vy\\[0.1em]
\big(\vh_4^{T}\vh_4\big)^{-1}\vh_4^{T}\ \vy
\end{bmatrix},
\end{eqnarray}
and $\vh_i$ is the $i$th column of channel matrix $\mh,\ i=1,2,3,4$.
The SM detector output $y(n)$ is fed to the $N$-point FFT. 
The subcarriers 1 to $\big(\frac{N}{2}-1\big)$ at the FFT output and
demodulated to get back the transmit data. The index bits $b(n)$s are 
detected as $\hat{b}(n)= 0$ if 
$\displaystyle\arg\max_{i=1,2,3,4} |z_i(n)|=1$ or 2.
$\hat{b}(n) = 1$ if 
$\displaystyle\arg\max_{i=1,2,3,4} |z_i(n)|=3$ or 4.

\begin{figure}
\centering
\includegraphics[scale=.6]{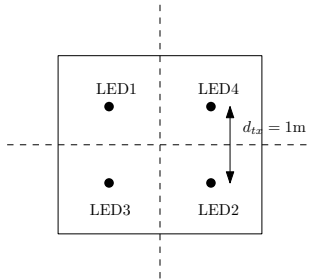}
\caption{Placement of $N_t=4$ LEDs in a $2\times 2$ grid with $d_{tx}=1$m.}
\label{indcplac}
\end{figure}

\subsection{Performance of I-NDC OFDM}
\label{sec3c}
Here, we present the BER  performance of the proposed I-NDC OFDM scheme 
for various system parameters. We fix the number of LEDs in I-NDC OFDM 
to be $N_t=4$ (see Fig. \ref{indctx}), and the number of PDs to be $N_r=4$. 
The placement of $N_t=4$ LEDs in a $2\times 2$ square grid is shown in 
Fig. \ref{indcplac}. We also compare the performance of the proposed
I-NDC OFDM with that of NDC OFDM. LED2 and LED3 are used for NDC OFDM.

\begin{figure}
\centering
\includegraphics[width=3.35in,height=2.5in]{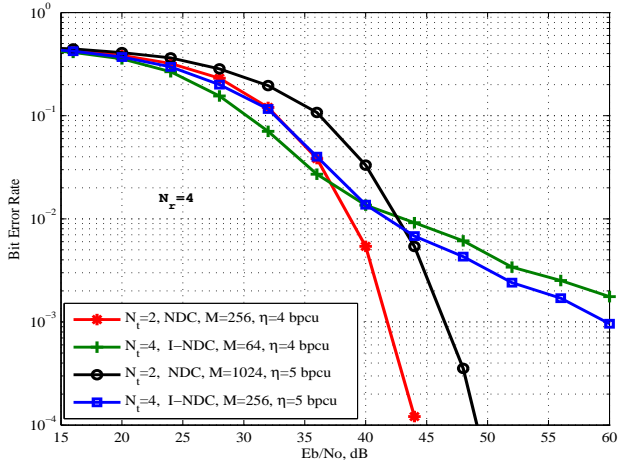}
\vspace{-2mm}
\caption{Comparison of the BER performance of NDC OFDM and the proposed
I-NDC OFDM for $\eta=4,\ 5$ bpcu, $N_r=4$.}
\label{fig5}
\vspace{-4mm}
\end{figure}

Figure \ref{fig5} presents the BER performance comparison of I-NDC OFDM 
and NDC OFDM for $\eta=4,\ 5$ bpcu. We fix the number of PDs to be $N_r=4$ 
for both I-NDC OFDM and NDC OFDM. The parameters considered in these 
systems for $\eta=4$ bpcu are: $1$) NDC OFDM: $N_t=2$, $N_r=4$, $M=256$, 
and $2$) I-NDC OFDM: $N_t=N_r=4$, $M=64$. Similarly, the system parameters 
considered for $\eta=5$ bpcu are: $1$) NDC OFDM: $N_t=2$, $N_r=4$, $M=1024$, 
and $2$) I-NDC OFDM: $N_t=N_r=4$, $M=256$. From Fig. \ref{fig5}, it can be 
seen that the I-NDC OFDM outperforms NDC OFDM at low SNRs. This is because,
to achieve the same spectral efficiency, I-NDC OFDM uses a smaller-sized QAM 
compared to that in NDC OFDM. But, as the SNR increases, the NDC OFDM 
outperforms I-NDC OFDM. This is because, as the number of LEDs is increased, 
the channel correlation increases which affects the detection performance.
Note that, though only one LED will be active at a time in both NDC OFDM 
as well as I-NDC OFDM, NDC OFDM has 2 LEDs whereas I-NDC OFDM has 4 LEDs. 

\begin{figure}
\centering
\includegraphics[width=3.35in,height=2.5in]{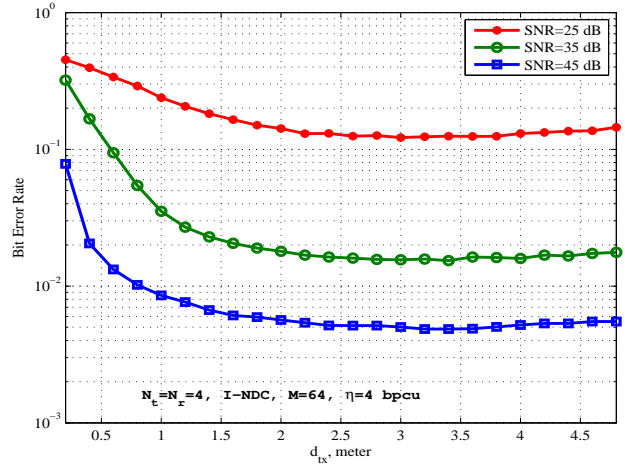}
\vspace{-2mm}
\caption{Comparison of the BER performance of I-NDC OFDM for varying 
$d_{tx}$, $\eta=4$ bpcu, and $N_t=N_r=4$.}
\label{fig6dtx}
\vspace{-4mm}
\end{figure}

In Fig. \ref{fig6dtx}, we present the BER performance of I-NDC OFDM as a
function of the spacing between the LEDs ($d_{tx}$) by fixing other system 
parameters. The parameters considered are: $N_t=N_r=4$, $M=64$ and $\eta=4$ 
bpcu, and SNRs = 25, 35, 45 dB. It is observed from Fig. \ref{fig6dtx} that 
there is an optimum $d_{tx}$ which achieves the best BER performance. The 
optimum spacing is found to be 3.4m in Fig. \ref{fig6dtx}. The BER 
performance get worse at $d_{tx}$ values those are above and below the 
optimum spacing.  This happens due to opposing effects of the channel 
gains and the channel correlations. That is, as $d_{tx}$ increases, the
channel correlation reduces and which improves the BER performance. On 
the other hand, the the channel gains get weaker as the $d_{tx}$ increases
and this degrades the BER. 

\begin{figure}[h]
\centering
\includegraphics[width=3.35in,height=2.5in]{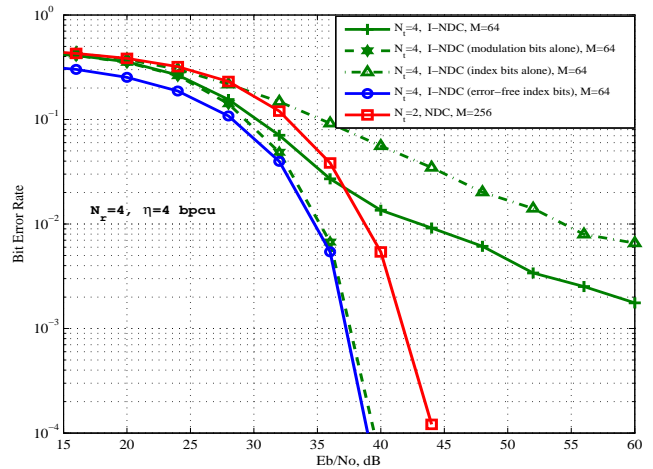}
\vspace{-2mm}
\caption{Reliability of modulation bits and index bits in the proposed
I-NDC OFDM for $\eta=4$ bpcu, $N_r=4$.}
\label{figx1}
\vspace{-4mm}
\end{figure}

\begin{figure*}
\centering
\includegraphics[width=5.5in]{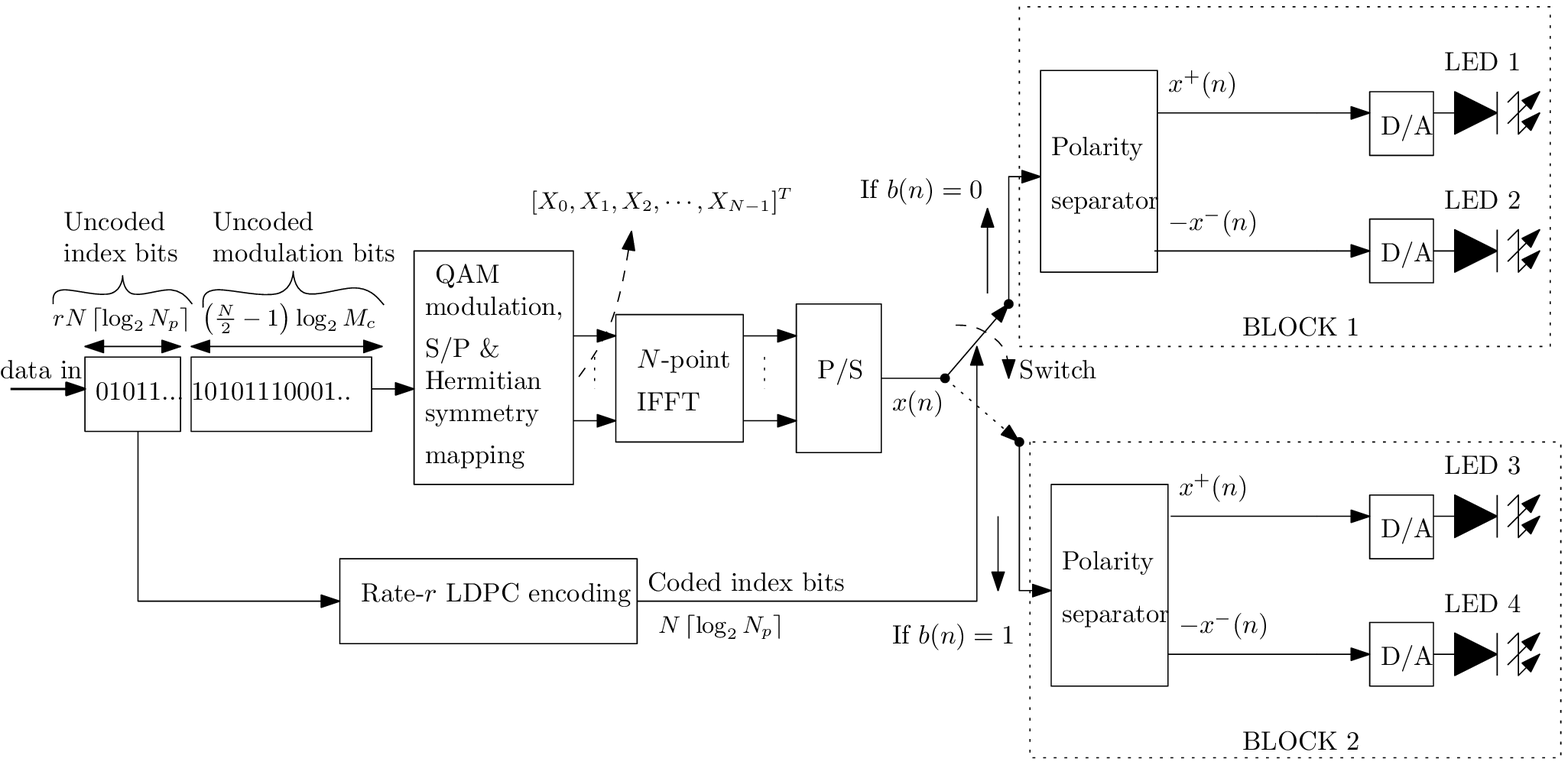}
\caption{Proposed CI-NDC OFDM transmitter.}
\label{cindctx}
\end{figure*}
\begin{figure*}
\centering
\includegraphics[width=4.75in]{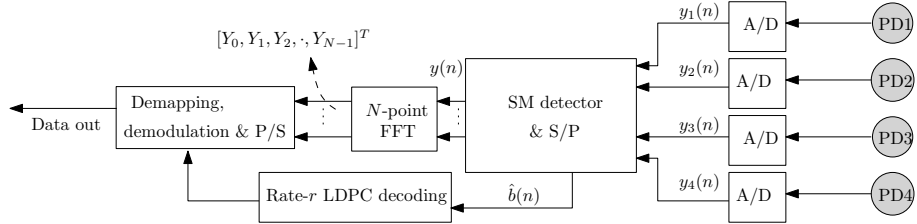}
\caption{Proposed CI-NDC OFDM receiver.}
\label{cindcrx}
\end{figure*}

\subsection{Proposed CI-NDC OFDM}
\label{sec3b}
{\em Motivation for CI-NDC OFDM:}
While investigating the poor performance of I-NDC OFDM at high SNRs, we
observed from the simulation results that reliability of the index bits 
is far inferior compared to the reliability of the modulation bits. This 
is illustrated in Fig. \ref{figx1}. As can be seen, the reliability of 
the index bits is so poor relative to the that of the modulation bits,
the overall performance is dominated by the performance of the index 
bits. This is because while the modulation bits have the benefit of OFDM 
signaling to achieve good performance, the index bits did not have any 
special physical layer care. This has motivated the need to provide some 
physical layer protection in the form of coding, diversity, etc. Indeed, 
as can be seen from Fig. \ref{figx1}, in the ideal case of error-free 
reception of index bits, the I-NDC OFDM has the potential of outperforming
NDC-OFDM even at high SNRs; see the plots of I-NDC OFDM (error-free 
index bits) and NDC OFDM. Motivated by this observation, we propose to use 
coding to improve the reliability of index bits.

{\em LDPC coding for index bits:} We propose to use a rate-$r$ LDPC code 
to encode $k$ uncoded index bits and obtain $n$ coded index bits, 
$r=\frac{k_c}{n_c}$. At the transmitter, $k_c$ uncoded index bits are 
accumulated to obtain $n_c$ LDPC coded index bits. Now, the $n_c$ coded 
index bits are used to select the index of the active LED block. Thus, 
one LDPC codeword of size $n_c$ is transmitted in 
$\frac{n_c}{\lfloor\log_2 N_p\rfloor}$ channel uses. Therefore, the 
overall spectral efficiency achieved by the CI-NDC scheme is
\begin{equation}
\eta_{cindc}=r\lfloor\log_2 N_p\rfloor+\frac{N-2}{2N}\ \log_2M_c\ \mbox{ bpcu},
\label{etacindc}
\end{equation}
where $M_c$ is the size of the QAM alphabet used in CI-NDC OFDM. 
The proposed CI-NDC OFDM transmitter and receiver are shown in Figs.
\ref{cindctx} and \ref{cindcrx}, respectively. 

\subsection{Performance of CI-NDC OFDM}
\label{sec3d}
\begin{figure}
\hspace{-2mm}
\includegraphics[width=3.75in,height=2.75in]{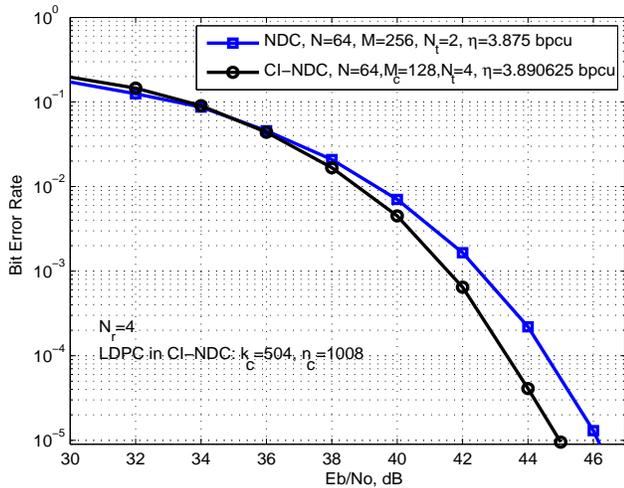}
\vspace{-6mm}
\caption{BER performance of the proposed CI-NDC OFDM and NDC OFDM 
at $\eta=3.8$ bpcu, $N_r=4$.}
\label{figx2}
\vspace{-4mm}
\end{figure}

In Fig. \ref{figx2}, we compare the performance of the proposed
C-INDC OFDM with that of NDC OFDM. We match the spectral efficiencies 
of both the schemes by using the following configurations: $1$) NDC 
OFDM: $N=64$, $M=256$, $N_t=2$, $N_r=4$, $\eta_{ndc}=3.875$ bpcu, and 
$2$) C-INDC OFDM: $N=64$, $M_c=128$, $N_t=4$, $N_r=4$, $r=\frac{1}{2}$, 
$k_c=504$, $n_c=1008$, $\eta_{cindc}=3.890625$ bpcu. From Fig. \ref{figx2}, 
we observe that, for the same spectral efficiency of about 3.8 bpcu, the 
proposed CI-NDC OFDM performs better than NDC OFDM. For example, to 
achieve a BER of $10^{-5}$, CI-NDC OFDM requires about 1.3 dB less SNR 
compared to NDC OFDM. This is because of the improved reliability of
the index bits achieved through coding of index bits. 

\section{Conclusions}
\label{sec5}
We proposed an efficient multiple LED OFDM scheme, termed
as coded index non-DC-biased OFDM, for VLC. The
proposed scheme was motivated by the high spectral efficiency
and performance benefits of using multiple LEDs and spatial
indexing. In the proposed scheme, additional information bits 
were conveyed through indexing in addition to QAM bits. The 
channel correlation in multiple LED settings was found to 
significantly degrade the reliability of index bits recovery. 
To overcome this, we proposed coding of index bits. This was 
found to serve the intended purpose of achieving better 
performance compared to other OFDM schemes for VLC.  
Investigation of the proposed signaling architecture for 
higher-order index modulation using multiple pairs of LEDs 
can be a topic of further study.

\end{document}